\documentclass[twocolumn]{jpsj3}
\usepackage{txfonts}
\usepackage{braket}
\usepackage{color}
\usepackage{bm}
\usepackage{graphicx}
\usepackage{epstopdf}

\usepackage{xcolor}

\title{Collinear Magnetic Structure in the Diamond Network Magnet EuTi$_2$Al$_{20}$}

\author{Masahiro Kawamata$^1$\thanks{kawamatam@tmu.ac.jp}, Ryuji Higashinaka$^1$, Takeshi Matsumura$^2$, Maxim Avdeev$^3$,\\
Kazuaki Iwasa$^{4,5}$, Hironori Nakao$^6$, Kazumasa Hattori$^1$, and Tatsuma D. Matsuda$^1$}
\inst{$^1$Department of Physics, Tokyo Metropolitan University, Hachioji, Tokyo 192-0397, Japan\\
$^2$Department of Quantum Matter, ADSE, Hiroshima University, Higashi-Hiroshima 739-8530, Japan\\
$^3$Australian Nuclear Science and Technology Organisation, Kirrawee DC, NSW 2232, Australia\\
$^4$Research and Education Center for Atomic Sciences, Ibaraki University, Tokai, Naka, Ibaraki 319-1106, Japan\\
$^5$Institute of Quantum Beam Science, Ibaraki University, Mito, Ibaraki 310-8512, Japan\\
$^6$Photon Factory, Institute of Materials Structure Science,
High Energy Accelerator Research Organization, Tsukuba, 305-0801, Japan
}

\abst{
The magnetic structure of EuTi$_2$Al$_{20}$, in which magnetic Eu$^{2+}$ ions form a diamond network, was investigated using neutron and resonant X-ray diffraction on powder and single-crystal samples.
The propagation vector was determined to be $\textbf{\textit{q}}_{\rm m}=(1,0,0)$~r.l.u. from these diffraction measurements.
All possible magnetic structures in the space group $Fd\bar{3}m$ with this propagation vector were examined using the irreducible representation method and magnetic space group analysis.
This magnetic structure was identified as a collinear antiferromagnetic structure with the magnetic space group $P_Inna$ (\#52.320) or $P_Inn2$ (\#34.164) under zero magnetic field.
In these magnetic structure, frustration arises from competing magnetic interactions on the diamond network. These findings provide a concrete experimental reference for assessing the role of competing interactions in diamond-network magnets and motivate further studies of interaction-driven quantum states.
}

\date{\today}
\begin{document}
\maketitle
\section{Introduction}
Many magnetic materials in nature achieve a stable ground state through cooperative ordering of spins. 
However, when the lattice geometry or competing interactions make it impossible to satisfy all pairwise exchanges simultaneously, conventional ordering can be impeded.
This situation, known as magnetic frustration, can prevent the system from establishing a unique ground state.
Research in magnetically frustrated systems has garnered considerable attention in recent years as a source of rich quantum phenomena in strongly correlated electron systems~\cite{Lacroix2011}. 
Representative examples include frustrated magnets on triangular~\cite{Starykh2015}, kagome~\cite{Mendels2016}, and pyrochlore systems~\cite{Castelnovo2012}. 
In these systems, strong degeneracy combined with quantum fluctuations often suppresses conventional long-range orders down to very low temperatures. 
As a result, non-trivial ground states such as quantum spin liquids and spin ice have been proposed and extensively studied both experimentally and theoretically for decades~\cite{Balents2010,Coldea2001,Nakatsuji2005,Mendels2010,Higashinaka2004}. 
In particular, the avoidance of ordering has attracted significant interest as hallmarks of novel physics beyond the conventional paradigm of magnetic orders.

\begin{figure}
    \includegraphics[width=1\linewidth]{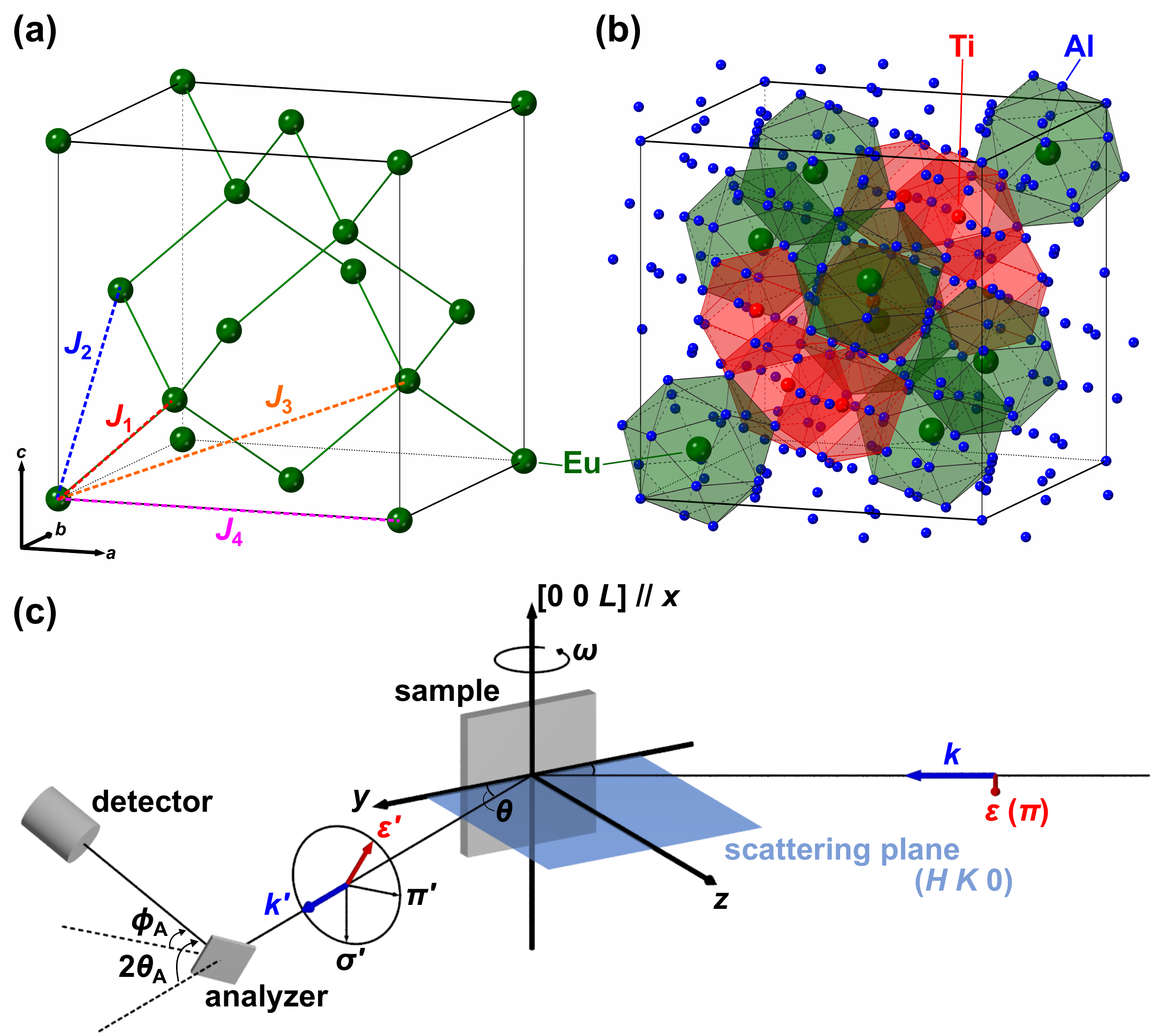}
    \caption{(a) The diamond network composed of Eu sites of EuTi$_2$Al$_{20}$. The first-, second-, third-, and fourth-neighbor exchange interaction terms are denoted by $J_1$, $J_2$, $J_3$, and $J_4$. (b) Crystal structure of EuTi$_2$Al$_{20}$ showing the network of the Eu-Al/Ti-Al cages. Note that it is translated by $(-1/8,-1/8,-1/8)$ relative to (a). (c) Configuration of resonant X-ray diffraction (RXD) experiment.}
    \label{f1}
\end{figure}

Beyond these canonical frustrated systems, frustration can also emerge on other three-dimensional networks through competing interactions.
For example, the diamond network, consisting of two interpenetrating face-centered cubic (fcc) lattices displaced by $(1/4,1/4,1/4)$, can host frustrated states driven by competing interactions.
Considering only the nearest-neighbor exchange interaction $J_1$ between the sublattices [Fig.~\ref{f1}(a)], an antiferromagnetic (AFM) $J_1$ stabilizes a N\'{e}el-type AFM order~\cite{Ge2018}.
When next-nearest-neighbor interactions $J_2$ are introduced, however, frustration can arise: irrespective of whether $J_1$ is ferromagnetic (FM) or AFM, and the simple N\'{e}el state is destabilized and stabilizes nontrivial magnetic orders emerge for $|J_2/J_1|\geq 1/8$ when the $J_2$ is antiferroic~\cite{Bergman2007}.
This scenario has been extensively studied in magnetic A-site spinels, where magnetic ions form a diamond network~\cite{Lee2008}.
Owing to the three-dimensional connectivity of this network, the propagation of quantum fluctuations and spin correlations differs significantly from that in low-dimensional systems. 
Reports of quantum-spin-liquid-like behavior and field-induced skyrmion phases in diamond-network magnets indicate that the diamond network is a fertile platform for frustration physics~\cite{Zaharko2011,Gao2017}.

Frustration in diamond networks can originate not only from competing short-range exchange interactions in insulating systems but also from long-range Ruderman-Kittel-Kasuya-Yosida (RKKY) interactions mediated by conduction electrons~\cite{Ruderman1954,Kasuya1956,Yosida1957}. 
A prominent family of materials exemplifying this scenario is the intermetallic $RT_2X_{20}$ compounds ($R$: rare-earth, $T$: transition metal, $X$: Al, Zn, and Cd), which crystallize in a cage-type structure with the $R$ ions forming a diamond network [Fig.~\ref{f1}(b)].
This structure belongs to the space group $Fd\bar{3}m$, with the $R$ ions occupying the $8a$ Wyckoff site.
Following the origin choice 2 of this space group, Fig.~\ref{f1}(b) is translated by $(-1/8,-1/8,-1/8)$ relative to (a).
In these systems, diverse ground states based on RKKY interactions have been observed. For instance, quadrupole order occurs in Pr-based systems~(e.g., Refs.~[\citen{Sakai2011,Onimaru2011,Onimaru2012}]), while heavy-fermion behavior has been reported in Yb- and Sm-based systems~\cite{Torikachvili2007,Higashinaka2011,Yamada2015,Afzal2024}. 
These properties highlight that the low-temperature physics is primarily governed by the $R$-ion multipolar degrees of freedom. 
Nevertheless, despite extensive investigations of multipole orders and correlated electron phenomena, the $RT_2X_{20}$ family has remained largely unexplored in the perspective of frustration magnetism arising from competing RKKY interactions in the diamond network.

In this work, we focus on EuTi$_2$Al$_{20}$, in which Eu$^{2+}$ ions form a diamond network. 
This compound exhibits an AFM transition at $T_{\rm N}=3.3$ K, with magnetism arising from Eu$^{2+}$ ($S=7/2, L=0$)~\cite{Kumar2016,Higashinaka2025}. 
The Eu moments couple via RKKY interactions, which are nearly Heisenberg-like due to the absence of orbital angular momentum. 
Magnetization measurements have revealed a field-induced intermediate phase characterized by a half-magnetization plateau, accompanied by enhanced magnetoresistance and unconventional Hall effect behavior~\cite{Higashinaka2025}. 
The Hall response cannot be explained solely by the sum of the normal and anomalous contributions, suggesting an additional mechanism, possibly related to emergent magnetic fields from exotic spin textures such as skyrmions~\cite{Tokura2021}.
Neutron diffraction experiments previously reported a propagation vector $\textbf{\textit{q}}_{\rm m} = (1, 0, 0)$ r.l.u.~\cite{Kurumaji2021}.
To elucidate the zero-field magnetic structure of EuTi$_2$Al$_{20}$, we have carried out neutron powder diffraction and resonant X-ray diffraction measurements on single-crystals.
Our results demonstrate the realization of a collinear AFM structure, at least in zero magnetic field.

\section{Experimental}
Single-crystals were grown by the Al self-flux method, following the procedure described in Ref.~\cite{Higashinaka2025}.
The samples were prepared for both neutron powder diffraction (NPD) and resonant X-ray diffraction (RXD) experiments.

The NPD experiment was carried out using the Echidna diffractometer~\cite{Avdeev2018} at the OPAL research reactor, ANSTO, Australia.
The incident neutron wavelength was set to 2.4431~\AA.
$\sim$1.0~g powder sample was sealed in a vanadium can and mounted in a cryostat.
Because Eu has a very large neutron absorption cross section ($\sigma_{\rm abs}=4530$~barn~\cite{Sears1992}), a small-diameter can ($\phi=2.7$~mm) was employed to reduce absorption.

The RXD experiment was performed at beamline BL-3A of the Photon Factory, KEK, Japan.
The experimental scattering geometry is illustrated in Fig.~\ref{f1}(c).
The X-ray energy was tuned to the vicinity of the Eu $L_2$ edge.
The incident beam was linearly polarized within the $y-z$ plane as of $\boldsymbol{\varepsilon} (\pi)$.
For polarization analysis, a pyrolytic graphite (PG) analyzer crystal with the (0 0 6) reflection ($2\theta_{\rm A}=93.7^\circ$ at the Eu $L_2$ edge) was employed.
The analyzer rotation angle, $\phi_{\rm A}$, was used to estimate the scattered X-ray polarization components $\sigma'$ and $\pi'$ of $\boldsymbol{\varepsilon}'$.
A single-crystal specimen was spark-cut into a plate shape with dimensions $1.7\times1.1$~mm$^2$ and a thickness of 0.5~mm.
The $(1\ 0\ 0)$ surface was polished to a mirror-like finish.
The crystal, oriented in the $(H\ K\ 0)$ horizontal scattering plane, was mounted in a cryostat.

\section{Results and discussion}
\begin{figure}
    \includegraphics[width=1\linewidth]{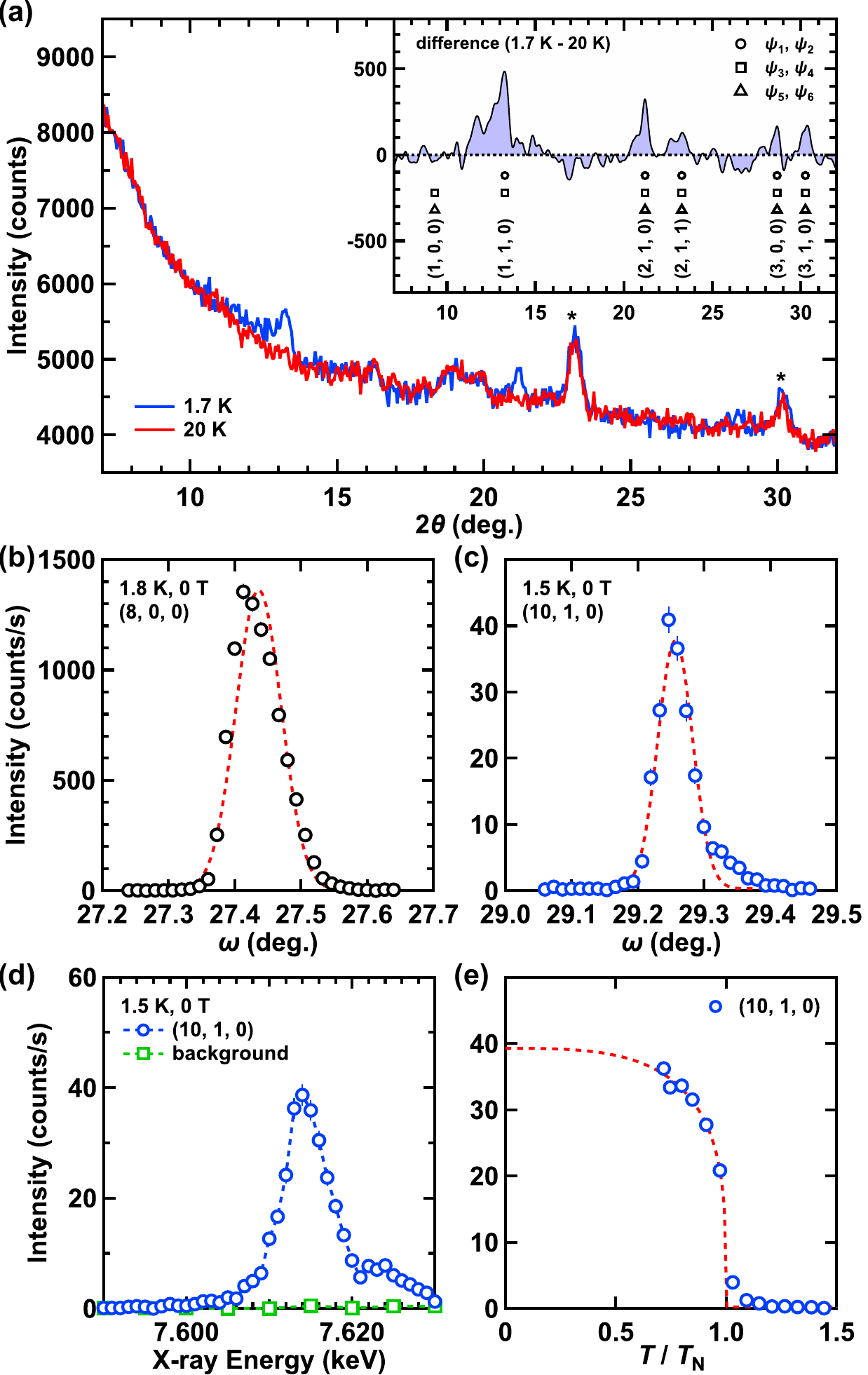}
     \caption{(a) Neutron powder diffraction (NPD) patterns at 1.7~K and 20~K. The reflection angles for each basis vectors (BVs) are indicated by black bars. The star marker indicates an unknown reflection. Inset: difference pattern (1.7~K - 20~K) with the expected positions of magnetic reflections from $\psi_1-\psi_6$.(b-e) Resonant X-ray diffraction (RXD) results. (b) Rocking curve (RC) of the fundamental reflection $(8,0,0)$ at 1.8~K and 0~T. (c) RC of the magnetic reflection $(10,1,0)$ at 1.5~K and 0~T. (d) Energy and (e) temperature dependence of the magnetic reflection $(10,1,0)$.}
    \label{f2}
\end{figure}

Figure~\ref{f2}(a) displays the raw NPD data collected at 1.7 and 20~K [blue and red line in Fig.~\ref{f2}(a)], representing temperatures below and above the magnetic transition, respectively.
The diffraction pattern at 1.7~K exhibits several additional reflections that are absent at 20~K.
The difference between the two diffraction patterns can be attributed to magnetic reflections with $\textbf{\textit{q}}_{\rm m}=(1,0,0)$ r.l.u. [inset in Fig.~\ref{f2}(a)], consistent with the previous studies~\cite{Kurumaji2021}.

To determine the magnetic structure, we performed representational analysis at the \textit{X} point $(1,0,0)$~r.l.u. based on the irreducible representations (irreps), using the Bilbao Crystallographic Server to obtain the irrep labels and the magnetic-representation decomposition~\cite{Aroyo2006}, and SARAh to generate the corresponding symmetry-adapted basis vectors (BVs)~\cite{Wills2000}.
Four magnetic irreps, labeled ${\rm mX}_1$-${\rm mX}_4$ following this labeling convention, are allowed by symmetry, and each irrep contains two BVs.
For Eu at the $8a$ site, the magnetic representation decomposes as ${\rm mX}_2 \oplus {\rm mX}_3 \oplus {\rm mX}_4$, as summarized in Table~\ref{t1}.
The BVs belonging to ${\rm mX}_2$ correspond to moments oriented parallel to $\textbf{\textit{q}}_{\rm m}$, whereas those of ${\rm mX}_3$ and ${\rm mX}_4$ include components perpendicular to both $\textbf{\textit{q}}_{\rm m}$ and the BVs of ${\rm mX}_2$.

BVs $\psi_3$-$\psi_6$ belonging to ${\rm mX}_3$ and ${\rm mX}_4$ are expected to generate strong magnetic reflections at the $(1,0,0)$ position ($2\theta \approx 9.5^\circ$) in the NPD pattern.
However, within the resolution of the present experiment, no such reflection was detected.
This absence indicates that the magnetic structure is not described by ${\rm mX}_3$ or ${\rm mX}_4$. Instead, we assign the magnetic structure to ${\rm mX}_2$, which is spanned by the BVs $\psi_1$ and $\psi_2$.
Because $\psi_1$ and $\psi_2$ have components parallel to $\textbf{\textit{q}}_{\rm m}$, the resulting magnetic configuration is collinear.
The NPD patterns produced by $\psi_1$ and $\psi_2$ are indistinguishable, and thus the relative weight of the two BVs cannot be determined from powder data alone.
A quantitative evaluation of the ordered moment at each site will require single-crystal neutron diffraction measurements over a wide $\textbf{\textit{Q}}$ range.

\begin{table}
     \caption{Basis vectors (BVs) of irreducible representations (irreps) for the space group $Fd\bar{3}m$ with the propagation vector $\textbf{\textit{q}}_{\rm m}=(1,0,0)$~r.l.u. The atoms are defined as \#1: $(1/8,1/8,1/8)$ and \#2: $(7/8,7/8,7/8)$.}
     \begin{tabular}{cc|ccc|ccc}
     \hline\hline
                    &              & \multicolumn{3}{c}{atom~\#1}   & \multicolumn{3}{c}{atom~\#2}   \\
     \hline
     irrep          & BV           & $m_x$   & $m_y$   & $m_z$   & $m_x$   & $m_y$   & $m_z$    \\
     \hline
     ${\rm mX_2}$     & $\psi_1$     & 8       & 0       & 0       & 0       & 0       & 0
     \\
     ${\rm mX_2}$     & $\psi_2$     & 0       & 0       & 0       &8       & 0       & 0 
     \\
     ${\rm mX_3}$     & $\psi_3$     & 0       & 4       & 0       & 0       & 0      & -4    
     \\
     ${\rm mX_3}$     & $\psi_4$     & 0       & 0       & 4       &  0      & -4       & 0 
     \\
     ${\rm mX_4}$     & $\psi_5$     & 0       & 4      & 0       &  0      & 0       & 4 
     \\
     ${\rm mX_4}$     & $\psi_6$     & 0       & 0       & -4       & 0       & -4       & 0 
     \\\hline
     \end{tabular}
     \label{t1}
\end{table}

Next, to further investigate the domain structure and to determine the magnetic structure in detail, we performed resonant X-ray diffraction (RXD) experiments using single-crystals.
These measurements revealed that a collinear AFM structure is realized at zero magnetic field.

Figure~\ref{f2}(b) presents the rocking curve (RC) of the fundamental Bragg reflection $(8,0,0)$.
A Gaussian fit yields a full width at half maximum (FWHM) of $0.0809(6)^\circ$, indicating a small mosaic spread and crystalline quality sufficient for RXD.
Figure~\ref{f2}(c) shows the RC at $(10,1,0)$.
The energy dependence of this reflection peaks at 7.614~keV, coincident with the Eu $L_2$ edge~[Fig.~\ref{f2}(d), blue marker].
At positions slightly offset from the magnetic Bragg point, the intensity is comparable to the background~[Fig.~\ref{f2}(d), green marker], implying that unwanted contributions such as fluorescence are negligible.
Moreover, this peak vanishes above $T_{\rm N}$ [Fig.~\ref{f2}(e)].
Taken together, these observations establish that the signal in Fig.~\ref{f2}(c) arises from resonant magnetic scattering.

\begin{figure}
    \includegraphics[width=1\linewidth]{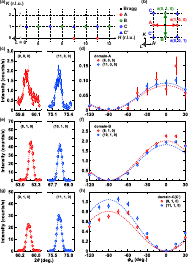}
    \caption{(a) Position of the observed magnetic reflection at 1.5~K and 0~T. (b) Relation between the fundamental Bragg reflection and the magnetic reflection. (c), (e), (g) $2\theta$ dependence of the magnetic reflection, and (d), (f), (h) corresponding polarization analysis of the magnetic reflection, respectively.}
    \label{f3}
\end{figure}

Figure~\ref{f3}(a) shows the positions of magnetic reflections observed in RXD on the $(H\ K\ 0)$ plane.
Based on the NPD results, the propagation vector is $\textbf{\textit{q}}_{\rm m}=(1,0,0)$ r.l.u., and the following magnetically equivalent domains exist crystallographically: $\textbf{\textit{q}}_{\rm A}=(1,0,0)$, $\textbf{\textit{q}}_{\rm B}=(0,1,0)$, and $\textbf{\textit{q}}_{\rm C}=(0,0,1)$ r.l.u.
Within $(H\ K\ 0)$ plane, magnetic reflections were detected at $\pm\textbf{\textit{q}}_{\rm A}$, $\pm\textbf{\textit{q}}_{\rm B}$, and $\pm\textbf{\textit{q}}_{\rm C}$ relative to the fundamental Bragg reflections.

To explain the observed positions of the magnetic reflections, we calculated the magnetic structure factor.
The magnetic scattering amplitude in resonant X-ray diffraction can be expressed as~\cite{Lovesey1996},
\begin{align}
f_{\rm m} &\propto (\boldsymbol{\varepsilon}^\prime \times \boldsymbol{\varepsilon})\cdot \textbf{\textit{F}}_{\rm m}(\textbf{\textit{Q}}),
\label{eq1}
\end{align}
where $\boldsymbol{\varepsilon}$ and $\boldsymbol{\varepsilon}^\prime$ are the polarization vectors of the incident and scattered X-rays, respectively, $\textbf{\textit{Q}}=\textbf{\textit{k}}^\prime-\textbf{\textit{k}}$ is the scattering vector, and $\textbf{\textit{F}}_{\rm m}(\textbf{\textit{Q}})$ is the resonant magnetic structure factor given by,
\begin{align}
\textbf{\textit{F}}_{\rm m}(\textbf{\textit{Q}}) &= \sum_i \textbf{\textit{m}}_i e^{-i\textbf{\textit{Q}}\cdot \textbf{\textit{r}}_i},
\label{eq2}
\end{align}
where the sum runs over the eight Eu ions in the unit cell.
Let the coefficients of the BVs $\psi_1$ and $\psi_2$ be $m_1$ and $m_2$, respectively.
The resonant magnetic structure factor $F_{\rm m}^{\textbf{\textit{q}}_{\rm m}}$ corresponding to each propagation vector $\textbf{\textit{q}}_{\rm m}$ can then be expressed as follows within the $(H\ K\ 0)$ plane.
\begin{align}
|\textbf{\textit{F}}_{\rm mX_2}^{\textbf{\textit{q}}_{\rm A}}(\textbf{\textit{Q}})|^2 &=
\left\{
\begin{array}{ll}
4\sqrt{m_1^2+m_2^2} & H; {\rm odd},\ K; {\rm even}\\
0& {\rm otherwise}
\end{array}
\right.\label{eq3}\\
|\textbf{\textit{F}}_{\rm mX_2}^{\textbf{\textit{q}}_{\rm B}}(\textbf{\textit{Q}})|^2 &=
\left\{
\begin{array}{ll}
4\sqrt{m_1^2+m_2^2} & H; {\rm even},\ K; {\rm odd}\\
0& {\rm otherwise}
\end{array}
\right.\label{eq4}\\
|\textbf{\textit{F}}_{\rm mX_2}^{\textbf{\textit{q}}_{\rm C}}(\textbf{\textit{Q}})|^2 &=
\left\{
\begin{array}{ll}
4(m_1+m_2) & H,K; {\rm odd},\ n\in\mathbb{Z},\ H+K=4n \\
4(m_1-m_2) & H,K; {\rm odd},\ n\in\mathbb{Z},\ H+K=4n+2\\
0& {\rm otherwise}
\end{array}
\right.\label{eq5}
\end{align}
The positions of the magnetic reflections are labeled A, B, C, and C$'$ [Fig.~\ref{f3}(b)].
The magnetic domains characterized by $\textbf{\textit{q}}_{\rm A}$ and $\textbf{\textit{q}}_{\rm B}$ give rise to reflections A and B, respectively~[Eqs.~(\ref{eq3}) and (\ref{eq4})], while the domain with $\textbf{\textit{q}}_{\rm C}$ produces reflections C and C$'$~[Eq.~(\ref{eq5})].
Figures~\ref{f3}(c), \ref{f3}(e), and \ref{f3}(g) show unpolarized-diffraction data collected without a PG analyzer crystal for each domain selection.
The finite intensities observed at A, B, C, and C$'$ are consistently accounted for by the magnetic structure model based on $\mathrm{mX}_2$~[Eqs.~(\ref{eq3})-(\ref{eq5})], once crystallographic-domain effects are taken into consideration.

Furthermore, we performed polarization analysis for each magnetic reflection to determine the orientation of the magnetic moments.
Figures~\ref{f3}(d), \ref{f3}(f), and \ref{f3}(h) show the analyzer scan for the reflections belonging to the $\textbf{\textit{q}}_{\rm A}$, $\textbf{\textit{q}}_{\rm B}$, and $\textbf{\textit{q}}_{\rm C}$ domains, respectively.
In the $\textbf{\textit{q}}_{\rm A}$ and $\textbf{\textit{q}}_{\rm B}$ domains, the intensity reaches a maximum at $\phi_{\rm A} = 0^\circ$, whereas in the $\textbf{\textit{q}}_{\rm C}$ domain it peaks at $\phi_{\rm A} = -90^\circ$.
According to Eq.~(\ref{eq1}), this polarization dependence originates from the cross products $(\boldsymbol{\varepsilon}_{\sigma^\prime}^\prime\times\boldsymbol{\varepsilon}_{\pi})$ and $(\boldsymbol{\varepsilon}_{\pi^\prime}^\prime\times\boldsymbol{\varepsilon}_{\pi})$, which, in our scattering geometry, read~\cite{Lovesey1996},
\begin{align}
\boldsymbol{\varepsilon}_{\sigma^\prime}^\prime\times\boldsymbol{\varepsilon}_{\pi} &= -\textbf{\textit{e}}_y\cos{\theta}+\textbf{\textit{e}}_z\sin{\theta},\\
\boldsymbol{\varepsilon}_{\pi^\prime}^\prime\times\boldsymbol{\varepsilon}_{\pi} &= -\textbf{\textit{e}}_x\sin{2\theta}.
\end{align}
Here $\theta$ is the Bragg angle and $\textbf{\textit{e}}_{x,y,z}$ are unit vectors of the laboratory frame [Fig.~\ref{f1}(c)].
The analyzer angle $\phi_{\rm A}$ selects the scattered polarization component, with $\phi_{\rm A}=0^\circ$ and $-90^\circ$ corresponding to $\sigma^\prime$ and $\pi^\prime$, respectively.
Crucially, the analyzer angle dependences $I(\phi_{\rm A})$ are captured quantitatively by fits of the form  $I(\phi_{\rm A})=A\cos^2{(\phi_{\rm A}-\phi_0)}$ derived from Eq.~(\ref{eq1}), yielding $\phi_{\rm 0}=0^\circ$ for $\textbf{\textit{q}}_{\rm A,B}$ [maxima at $\phi_{\rm A}=0^\circ$ in Figs.~\ref{f3}(d), \ref{f3}(f)] and $\phi_{\rm 0}=-90^\circ$ for $\textbf{\textit{q}}_{\rm C}$ [maxima at $\phi_{\rm A}=-90^\circ$ in Fig.~\ref{f3}(h)], thereby validating the assigned moment directions. 
Accordingly, the $\textbf{\textit{q}}_{\rm A}$ and $\textbf{\textit{q}}_{\rm B}$ domains host moments confined to the $ab$ plane, while the $\textbf{\textit{q}}_{\rm C}$ domain has moments along $c$.
These observations indicate $\textbf{\textit{m}} \parallel \textbf{\textit{q}}_{\rm m}$, consistent with the irreducible representation ${\rm mX}_2$.

\begin{figure}
    \includegraphics[width=1\linewidth]{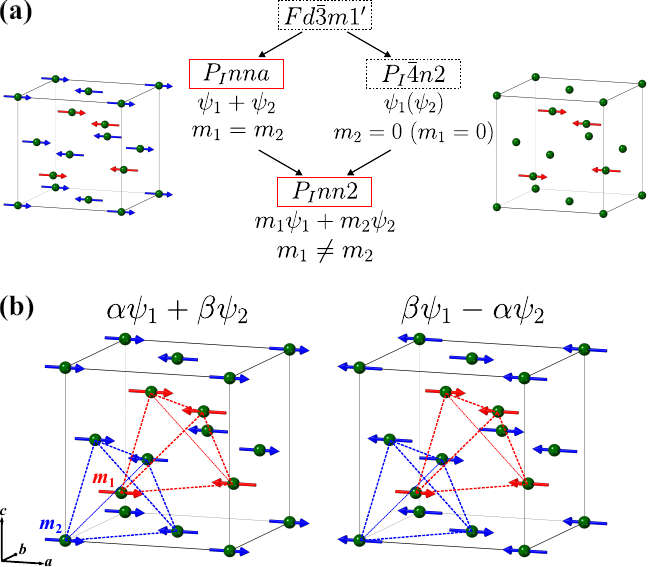}
    \caption{(a) Magnetic subgroup corresponding to the irreducible representation mX$_2$. (b) Magnetic structure of $P_Inn2$ (\#34.164). Left: $(m_1,m_2)=(\alpha,\beta)$, Right: $(m_1,m_2)=(\beta,-\alpha)$ with $\textbf{\textit{q}}_{\rm A}=(1,0,0)$~r.l.u. The red and blue dashed lines trace the tetrahedral units on each sublattice of the diamond network. The red and blue arrows represent the magnetic moments at the Eu sites \#1 and \#2, respectively.}
    \label{f4}
\end{figure}

\begin{figure}
    \includegraphics[width=1\linewidth]{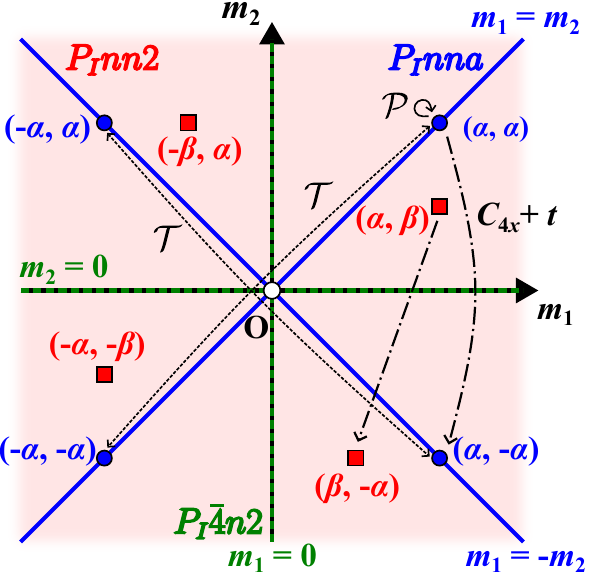}
    \caption{Order-parameter space of $(m_1,m_2)$ in two dimentional irrep ${\rm mX}_2$ with $\textbf{\textit{q}}_{\rm A}=(1,0,0)$~r.l.u. Blue solid and green dashed lines denote the $P_Inna$ and $P_I\bar{4}n2$, respectively. The red-shaded region represents $P_Inn2$. Symmetry operations are indicated by arrows: spatial inversion $\mathcal{P}$ with black solid arrows, time-reversal $\mathcal{T}$ with black dotted arrows, and $C_{4x} +\textbf{\textit{t}}$ ($\textbf{\textit{t}}=(-1/4,-1/4,-1/4)$) around Eu atom \#1 denoted in Table~\ref{t1} of the \textit{X}-point subgroup of $Fd\bar{3}m$ with black dash-dotted arrow.}
    \label{f5}
\end{figure}

We consider the magnetic space groups corresponding to the irreducible representation ${\rm mX}_2$.
Figure~\ref{f4}(a) shows the tree of possible subgroups within this irrep.
${\rm mX}_2$ is a two-dimensional representation using basis vectors $\psi_1$ and $\psi_2$. Figure \ref{f4}(b) shows the correspondence between the parameter space of the coefficients $(m_1, m_2)$ and the magnetic space group.
First, the maximal subgroup $P_Inna$ (\#52.320) corresponds to $m_1=\pm m_2$ and is described by $\psi_1 \pm \psi_2$ [on the blue solid line in Fig.~\ref{f5}].
Within this structure, the spatial inversion symmetry $\mathcal{P}$ [black solid arrow in Fig.~\ref{f5}] is preserved.
Note that time reversal symmetry $\mathcal{T}$ reverses the signs of $(m_1, m_2)$ [black dashed arrow in Fig.~\ref{f5}].
Next, the other maximal subgroup, $P_I\bar{4}n2$ (\#118.314), corresponds to $m_2=0$ (or $m_1=0$) and is described by $\psi_1$ (or $\psi_2$) alone, respectively [on the green dashed line in Fig.~\ref{f5}].
In $P_I\bar{4}n2$, however, the arrangement of magnetic moments is not allowed at one of the magnetic sites.
Two sublattice sites on the diamond network of Eu ions are connected by symmetry in the paramagnetic phase, making it unlikely that only one site would order first.
A lower-symmetry subgroup, $P_Inn2$ (\#34.164), can be constructed by taking arbitrary linear combinations, $m_1\psi_1+m_2\psi_2$ ($m_1\neq \pm m_2$) [on red area in Fig.~\ref{f5}].
For general $(m_1,m_2)=(\alpha,\beta)$ (excluding $\alpha=0$, $\beta=0$, $\alpha=\pm\beta$), the $\mathcal{P}$ and $\mathcal{T}$ are not preserved.

According to Eq.~(\ref{eq5}), when $m_1 \approx m_2$ (respectively $m_1 \approx -m_2$), the intensity at the C (respectively C$'$) position in Fig.~\ref{f3}(a) is expected to dominate over that of the conjugate reflection, which is strongly suppressed or even absent.
By contrast, there is no clear tendency for only one of the C or C$'$ reflections [(11,1,0) and (9,1,0)] to be strongly suppressed.
This behavior can be explained by the fact that the Seitz operation $\left\{C_{4x}\mid \bar{\tfrac{1}{4}},\bar{\tfrac{1}{4}},\bar{\tfrac{1}{4}}\right\}$ in the subgroup at the \textit{X} point of the paramagnetic $Fd\bar{3}m$ maps the crystallographic domain $(\alpha,\beta)$ to $(\beta,-\alpha)$ [dash-dotted arrow in Fig.~\ref{f5}].
Four crystallographic/magnetic domains are allowed in total with $\boldsymbol{q}_{\rm A}=(1,0,0)$, $\pm(\alpha,\beta)$ and $\pm(\beta,-\alpha)$ [red square markers in Fig.~\ref{f5}]; $(\alpha,\beta)$ and $(\beta,-\alpha)$ are illustrated in Fig.~\ref{f4}(b).
In the limit $\beta \to \alpha$, four crystallographic domains, $\pm(\alpha,\alpha)$ and $\pm(\alpha,-\alpha)$, are realized [blue circle markers in Fig.~\ref{f5}].

From Eq.~(\ref{eq5}), domains with $\pm(\alpha,\beta)$ yield $I^{\rm C}(\textbf{\textit{Q}})\propto(\alpha+\beta)^2$ and $I^{\rm C^{\prime}}(\textbf{\textit{Q}})\propto(\alpha-\beta)^2$, whereas domains with $\pm(\beta,-\alpha)$ give the opposite weighting, $I^{\rm C}(\textbf{\textit{Q}})\propto(\alpha-\beta)^2$ and $I^{\rm C^{\prime}}(\textbf{\textit{Q}})\propto(\alpha+\beta)^2$.
A spatial mixture of these domains therefore renders the average structure factors at C and C$'$ approximately equal, consistent with experiment.
We thus conclude that $P_Inna$ or $P_Inn2$ provides the best description of the zero-field magnetic structure of EuTi$_2$Al$_{20}$.

In the previous study, the effective magnetic moment is reported 6.77~$\mu_{\rm B}$/Eu, which is slightly smaller than the theoretical value of 7.94~$\mu_{\rm B}$/Eu for Eu$^{2+}$~\cite{Kumar2016, Higashinaka2025}.
This indicates that magnetic moment shrinkage exists even when Eu sites are fully ordered.
Our previous study detected no defects at the Eu-site~\cite{Higashinaka2025}, suggesting that the average valence might be slightly reduced at one or both magnetic sites.
Indeed, the weak peak on the high-energy side in Fig.~\ref{f2}(d) may indicate a minor Eu$^{3+}$ component~\cite{Chen2023} or XANES vibrations of Eu$^{2+}$.
If this shrinkage is uniform, then $\alpha=\pm\beta$, resulting in the highly symmetric $P_Inna$ phase.
Conversely, in the case of $\alpha\neq\beta$, it breaks spatial inversion symmetry and realizes the $P_Inn2$ phase.
To distinguish these, it is necessary to examine the valence at each site, requiring studies using local techniques such as Mössbauer spectroscopy~\cite{Koldemir2023,Pottgen2023} or NMR.
Also, a quantitative determination of the valence ($\alpha$ and $\beta$) will require the XAFS or diffraction study on a single domain, which we leave for our future work.

A comparison within the $RT_2X_{20}$ family clarifies the distinctiveness of EuTi$_2$Al$_{20}$.
Pr$T_2X_{20}$ ($T=$ Ti, V, Nb, Rh, Ta, and Ir; $X=$ Zn and Al) possesses a nonmagnetic $\Gamma_3$ doublet crystal-field ground state and exhibits quadrupole order~\cite{Sakai2011, Matsunami2011, Onimaru2011, Higashinaka2011Pr, Onimaru2012, Matsubayashi2012, Tsujimoto2014, Tokunaga2013, Onimaru2016, Taniguchi2016,Higashinaka2017, Sakai2025}.
In particular, in PrIr$_2$Zn$_{20}$ an applied magnetic field induces antiferromagnetic order with $\boldsymbol{q}_{\rm m}=(\tfrac{1}{2},\tfrac{1}{2},\tfrac{1}{2})$ r.l.u.~\cite{Iwasa2017}.

NdRh$_2$Zn$_{20}$, GdCo$_2$Zn$_{20}$, and TbCo$_2$Zn$_{20}$ realize dipolar AFM order at $\boldsymbol{q}_{\rm m}=(\tfrac{1}{2},\tfrac{1}{2},\tfrac{1}{2})$,~\cite{Yamamoto2022,Mardegan2016,Tian2010} whereas SmTi$_2$Al$_{20}$ with a field-insensitive heavy-fermion behavior exhibits a simple N\'eel-type order at $\boldsymbol{q}_{\rm m}=(0,0,0)$ r.l.u.~\cite{Higashinaka2025_Sm}
In contrast, EuTi$_2$Al$_{20}$ orders at the \textit{X} point, $\boldsymbol{q}_{\rm m}=(1,0,0)$ r.l.u., indicating a different balance and range of exchange interactions in the Eu compound.

From a theoretical viewpoint, ordering near the \textit{X} point on the diamond network requires $\lvert J_2/J_1\rvert \ge 2/3$, and a simple $J_1$\,\textendash\,$J_2$ model alone does not fully stabilize the \textit{X}-point order.~\cite{Bergman2007}
Inclusion of a ferromagnetic fourth-nearest neighbor interaction $J_4$ [Fig.~\ref{f1}(b)], which corresponds to the second-nearest neighbor on each fcc sublattice, is necessary to select $\boldsymbol{q}_{\rm m}=(1,0,0)$.~\cite{Ignatenko2008,Balla2020}
In EuTi$_2$Al$_{20}$, where the Eu$^{2+}$ moments couple via strongly distance-dependent RKKY interactions, such longer-range couplings arise naturally and can provide an efficient route to \textit{X}-point stabilization.
Although direct experimental determination of $J_{ij}(\boldsymbol{r})$ is challenging, a combined program of \textit{ab initio} calculations and inelastic neutron scattering on the spin-wave spectrum offers a practical path to quantify the effective spin Hamiltonian.~\cite{Paddison2024}
Moreover, the observation of collinear magnetic order suggests that higher-order spin-spin interactions are operative such as biquadratic terms~\cite{Singh2017}. 
Accordingly, it is necessary to move beyond the simple $J_1$-$J_2$ description commonly invoked for diamond networks and consider models that incorporate both long-range and higher-order couplings for EuTi$_2$Al$_{20}$ .
In this system, what appears as strong frustration within the $J_1$-$J_2$ limit more plausibly reflects competition among long-range and higher-order interactions.

It is also instructive to contrast EuTi$_2$Al$_{20}$ with other diamond-network magnets beyond the $RT_2X_{20}$ family. Magnetic A-site spinels provide representative cases: CoRh$_2$O$_4$ stabilizes a N\'eel state~\cite{Ge2018}, CoAl$_2$O$_4$ lies near the $J_2/J_1 \!\sim\! 1/8$ boundary with sample-dependent ground states~\cite{MacDougall2011}, and MnSc$_2$S$_4$ evolves from a spin-liquid regime into helical orders at lower temperatures~\cite{Gao2017}.
Furthermore, field-induced skyrmion phases with thermal Hall responses have been reported in this class of materials~\cite{Gao2020,Takeda2024}.
These insulating systems are governed primarily by short-range exchange, whereas EuTi$_2$Al$_{20}$ features competition among long-range, oscillatory RKKY interactions, providing an alternative route to frustration on common systems with diamond network.


A complementary perspective comes from intermetallics that also host a diamond network: the Laves-phase $R$Al$_2$ (C15) family, where the $R$ sublattice forms a diamond network.
Most members are FM,\cite{Nereson1966,Nereson1968,Lee1981,Leson1986} while EuAl$_2$ orders antiferromagnetically at $\boldsymbol{q}_{\rm m}=(0,0,0)$.\cite{Ouladdiaf1997}
This comparison highlights that the lattice motif alone does not determine the ordering wave vector; rather, the range and sign pattern of exchange couplings (e.g., RKKY) play a decisive role.

Within $RT_2X_{20}$, AFM order is realized for $R\neq \text{Eu}$ as well, yet ordering at the \textit{X} point is, to date, unique to EuTi$_2$Al$_{20}$.
This tendency is consistent with the broader trend that Eu, particularly Eu$^{2+}$, often shows magnetic behavior distinct from other rare-earth ions~\cite{Onuki2020}.

Finally, EuTi$_2$Al$_{20}$ exhibits an intermediate, field-induced phase characterized by a half-magnetization plateau and anomalous transport signatures.~\cite{Higashinaka2025}
These observations suggest a characteristic field-induced magnetic texture emerging from the competition among frustrated RKKY interactions under external fields.
Determining the magnetic structure in this regime remains an important challenge. Small-angle and polarized neutron scattering, angle-resolved resonant X-ray diffraction, and systematic decomposition of the Hall response (ordinary, anomalous, and possible topological components) would be particularly informative.
Together with quantitative exchange modeling, such studies will help establish the effective spin Hamiltonian underlying the \textit{X}-point order and its field evolution in EuTi$_2$Al$_{20}$.

\section{Conclusions}
In this study, we investigated the zero-field magnetic structure of the diamond network compound EuTi$_2$Al$_{20}$ using neutron powder diffraction and resonant X-ray diffraction.
Both techniques consistently revealed the stabilization of a uniaxial magnetic structure with $\textbf{\textit{m}}\parallel\textbf{\textit{q}}_{\rm m}$.
Polarization analysis of single-crystal RXD further demonstrated the presence of twelave energetically multi magnetic domains with $\textbf{\textit{q}}_{\rm m}=(1,0,0)$ r.l.u.
The ratio of magnetic moment lengths in each sublattice remains unclear, but it exhibits collinear antiferromagnetism.
Such a structure cannot be captured by a simple $J_1$-$J_2$ model, implying that longer-range RKKY interactions, involving $J_4$ or beyond, play an important role.
EuTi$_2$Al$_{20}$ thus represents a rare example of a metal that hosts a frustrated magnetic structure on a diamond network.
This places it as a promising platform for realizing complex spin textures arising from competition with applied magnetic fields.
To establish this possibility, a crucial next step is to determine the magnetic structure of EuTi$_2$Al$_{20}$ under magnetic fields.
\begin{acknowledgments}
We thank Yoshichika \={O}nuki, Yuji Aoki, and Hiroaki Kusunose for their valuable discussions.
This work was supported by the JSPS (Nos.~JP22K03517, JP22K03522, JP23K03332, JP23H04866, JP23H04867, JP23H04869, JP23H04870, JP24K00574, JP25K07228).
Work at ANSTO was performed under the user program (No.~18101).
Work at Photon Factory was performed under the user program (No.~2024S2-002).
\end{acknowledgments}
%

\end{document}